\documentclass{elsart}
\usepackage{amsmath,amsfonts,amssymb,graphicx}
\usepackage{mathrsfs,bbm}				

\usepackage[sort&compress]{natbib}			
\bibpunct{[}{]}{,}{n}{}{;}				

	\newcommand{\E}[1]{\ensuremath{\mathbb{E}\!\left[#1\right]}}

	\newcommand{\norm}[1]{\ensuremath{\|#1\|}}

	\newcommand{\abs}[1]{\ensuremath{\left|#1\right|}}

	\DeclareMathOperator{\Var}{Var}

	\newcommand{\vect}{\ensuremath{\textbf{r}}}
	
	\newcommand{\mean}[1]{\langle #1 \rangle}

	\newcommand{\eZ}{\check{\mathbf{e}}_{z}}
	
	\newcommand{\brho}{\boldsymbol{\rho}}

	\renewcommand{\Re}{\ensuremath{\mathbb{R}}}

	\DeclareMathAlphabet{\mathpzc}{OT1}{pzc}{m}{it}
\begin{document}
\begin{frontmatter}
\title{A fractional Brownian motion model for the turbulent refractive index in lightwave propagation.}
\author{Dar\'{\i}o G. P\'erez\corauthref{cor}},
\corauth[cor]{Corresponding author.}
\ead{dariop@ciop.unlp.edu.ar}
and
\author{Luciano Zunino}
\ead{lucianoz@ciop.unlp.edu.ar}
\author{Mario Garavaglia}
\ead{garavagliam@ciop.unlp.edu.ar}

\address{Centro de Investigaciones \'Opticas (CIOp), CC. 124 Correo Central,1900 La Plata, Argentina.}

\address{Departamento de F\'{\i}sica,
                 Facultad de Ciencias Exactas,
                 Universidad Nacional de La Plata (UNLP),
                 1900 La Plata, Argentina.}


\begin{abstract} 
It is discussed the limitations of the widely used markovian approximation applied to model the turbulent refractive index in lightwave propagation. 
It is  well-known the index is a passive scalar field. Thus, the actual knowledge about these quantities is used to propose an alternative stochastic process to the markovian approximation: the \textit{fractional Brownian motion.} This generalizes the former introducing memory; that is, there is correlation along the propagation path.
\end{abstract}

\end{frontmatter}

\section{Introduction}
Whenever a light beam propagates through the turbulent atmosphere, it experiments deflections due to  fluctuations in the refractive index. As a result of phase changes, the beam suffers displacements perpendicular to the original direction of propagation. This phenomenon is known as \textit{beam wandering, beam steering} or \textit{spot dancing}. The wandering is usually characterized in terms of its variance. Several authors have experimentally and theoretically treated this problem, or the equivalent problem of angular fluctuations, using different approaches.

The earliest studies of the problem of beam wandering were within the Geometric Optics approximation. Chernov \cite{book:chernov60} treated the ray-light propagation in a random medium as a continuous Markov process. This assumption enabled him to formulate a Fokker-Planck equation. Later, Beckmann \cite{paper:beckman65} used the ray equation to calculate the wander of a single ray. Both authors obtained similar formulas for the variance of the transverse displacements, $\mathbf{Q}=(Q_x,Q_y)$, observed after the light has  propagated a distance $L$. These formulas differ in a numerical factor, but the power-law
\begin{equation}
\Var{\mathbf{Q}}\propto L^3
\label{eq:var-L3}
\end{equation}
is found in both. 

However, Chernov and Beckmann introduced an arbitrary Gaussian covariance function for the refractive index fluctuation which is not physically plausible; also, these techniques do not include the effects of a finite beam diameter. That is, small-scale variations of the refractive index contribute to the beam spread but have little effect on the motion of the beam centroid. Therefore, Geometric Optics approximation tends to overestimate the magnitude of the wandering. Chiba \cite{paper:chiba71} tried to overcome this limitation. He assumed that only changes of the refractive index larger than beam diameter contribute to the wandering. Using a Kolmogorov-like structure function he also found a power-law dependence as in Eq.~\eqref{eq:var-L3}.

On the other hand, whenever the width of the beam is smaller than the turbulence inner scale the  approximations obtained from Geometric Optics are enough. These beams known as thin-beam were studied by Consortini and O'Donnell \cite{paper:consortini-odonell91}. Following the Beckmann paper, they analized experimentally and theoretically the dependence of thin-beam displacements with the propagation distance. They showed that, independently of the turbulence spectrum, the displacements variance grows like the third-power of the path length. That is, provided the fluctuations are small and
the propagation length $L$ is large compared against the outer scale $L_0$. In another paper, Consortini \etal, \cite{paper:consortini97} investigated experimentally, for laboratory-generated turbulence, the dependence of beam variance on propagation length in the case of strong turbulence. They also found a rough estimate of the third-power law dependence.

It must be stressed here that all the previous works are valid in the Geometric Optics limit (see Ref.~\cite{book:tatarskii61}, p. 120). That is,
\begin{equation*}
l_0 \gg \sqrt{L\lambda},
\end{equation*}
where $\lambda$ is the wavelength and $l_0$ the dimension of the inner scale.

Klyatskin and Kon \cite{paper:klyatskin72} described the propagation of light considering a scalar parabolic equation and assuming a markovian approximation for the atmospheric refractive index fluctuation,  $\epsilon(\brho;z)$---$\brho$ represents transversal coordinates $(x,y)$. Using this approximation they derived an equation for the angular beam wandering, $\boldsymbol{\theta}=d \mathbf{Q} /d z$. They found a dependence proportional to $L$ in accordance with the results above---the displacement and angular beam wandering variances are proportional with factor $L^2$ (see Ref. \cite{paper:chiba71}, Eq. (15)).

When using the markovian approximation the original covariance function $R$, obtained from the original structure function for the index, is replaced by an \textit{effective covariance} $A$ through: 
\begin{equation}
\E{\,\epsilon(\brho,z)\epsilon(\brho',z')}=\delta(z-z') A(\brho-\brho'),
\label{eq:markovian-cov}
\end{equation}
where 
\begin{equation*}
A(\brho)=\int_{\Re} R(\brho,z)\,dz
\end{equation*}
is a differentiable function, and $\E{\cdot}$ is taken with a gaussian probability measure. This means that the values of the index in the region $z>z'$ do not affect those at the point $(\brho',z')$---which eliminates backscattering and scattering at sufficiently large angles. This property known as ``dynamical causality'' (see Ref.~\cite{paper:tatarskii80}, p. 214) comes from the martingale property of the markovian approximation.\addtocounter{footnote}{1}\footnote{
Formally, it is said that a process $X_t$ is markovian or possesses the \textit{Markov property} if the future behavior of it given what has happened up to time $t$ is the same as the behavior obtained when starting the process at $X_t$---a detailed description can be found in Ref.~\cite{book:shiryayev84}.
}

Tatarsk\u{\i} and Zavorotny \cite{paper:tatarskii80} derived the conditions of validity for the markovian approximation. They found that it is applicable if all the characteristic dimensions arising from the wave propagation problem are small compared to the path length. As Ostoja-Starzewski explained \cite{paper:ostoja2001} an intuitive justification for the Markov property is that the ray-light on a long distance behaves as if it has suffered many independent refractions. Then, this approach holds only for long-path propagation. 

In Sec. \ref{sec:markov} the equivalence between the markovian approximation and the procedures usually followed after the use of the Geometric Optics approximation, referred above, is shown. Thus, the former model is only valid for long-path propagation and weak turbulence, or just strong turbulence.

Recently, it was shown that \textit{fractional Brownian motion processes} (fBm) could be used to describe the turbulent refractive index fluctuation \cite{book:th-dario2003} to model ray-light propagation. This is not new since these fractal stochastic processes have also been used to identify turbulence degraded wave-fronts \cite{paper:schwartz-baum94, paper:perezzunino2004a}. In Sec. \ref{sec:fbm} we will show that these processes match the requirements for passive scalar fields; that is, they are nearly Gaussian, have stationary increments, and obey the \textit{Kolmogorov-Obukhov-Corrsin (KOC) structure function}---the extension to passive scalars of the well-known Kolmogorov structure function. Moreover, it was verified \cite{paper:zunino2004}, through Wavelet Analysis of experimental data, that the wandering of a laser beam presents memory---for short distances. This is in accordance with the fBm model presented above. Otherwise, the markovian approximation is memoryless; therefore, it can not represent the behavior of the refractive index in all spatial scales. Under these circumstances the fBm is a good candidate to replace and extend the markovian approximation.

\section{Markovian approximation}\label{sec:markov}

Most of the past and present research in turbulent lightwave propagation, we have seen, is based directly or indirectly on the markovian approximation. Usually, when approximations from the Geometric Optics are used the markovian property is indirectly applied \cite{book:chernov60,paper:beckman65,paper:chiba71,paper:consortini-odonell91}. These papers propose a simple model: a ray-light beam (laser beam) propagates through a turbulent flow a distance $L$ and the position on a screen is evaluated.

The Geometric Optics ray-equation under the condition $\epsilon\ll 1$, small index pertubation, is
\begin{equation}
\frac{d^2 \mathbf{Q}}{dz^2} = \nabla_{Q} \epsilon(\mathbf{Q},z).
\label{eq:small-dis}
\end{equation}
The assumption leads to two independent differential equations; therefore, it is enough here to consider just one axis. Usually, what the literature refers as \textit{Geometric Optics approximation} is obtained taking the zeroth order approximation for the transverse displacement, $Q_x$, from the above equation. That is,
\begin{equation}
\Var{ Q_x} = \int^L_0 \int^L_0 \int^z_0\int^{z'}_0 \frac{\partial^2 B_{\epsilon}}{\partial x \partial x'}(\brho,s;\brho,s') d s\, ds' dz\, dz', 
\label{eq:variance-distribution}
 \end{equation}
where $B_{\epsilon} (\vect,\vect')$ denotes the covariance function of the index fluctuation at points $\vect=(\brho,s)$ and $\vect' =(\brho' ,s')$. Observe that to obtain this equation it should be assumed that the gradient of the turbulent index is bounded.

It is always supposed the turbulence is homogeneous and isotropic. Under these hypotheses the covariance of the index fluctuation, $B_{\epsilon}$, is assumed to be an even function of the coordinate difference:
\begin{equation}
B_{\epsilon}(\vect,\vect')= B_{\epsilon}(\norm{\vect - \vect'}),
\label{eq:non-mark-cov}
\end{equation}
where $\vect=(\brho,z)$, $\vect'=(\brho',z')$ and, as usual, $\norm{\cdot}$ is the norm. After some elaborated calculations the following result is obtained
\begin{equation}
\Var{ Q_x} =- \int^L_0 f(L,z) \frac{\partial^2 B_{\epsilon}}{\partial^2 x}(0,z)\, dz.
\label{eq:variance-final}
\end{equation}
where $f(L,z)=(2 L^3 - 3 L^2 z + z^3)/3$ is called \textit{filter function}. All the papers considered that the covariance has only significant values for $z\ll L$. Therefore, the second and third terms in the filter function can be neglected changing the latter equation to:
\begin{equation}
\Var{ Q_x} \simeq - \frac{2L^3}{3} \int^L_0 \frac{\partial^2 B_{\epsilon}}{\partial^2 x}(0,z)\, dz \simeq - \frac{2L^3}{3} \int^{\infty}_0 \frac{\partial^2 B_{\epsilon}}{\partial^2 x}(0,z)\, dz.
\label{eq:longpath}\end{equation}
This result holds for $L\gg L_0$, long-path propagation.

Approximations involved in the previous relation are equivalent to the markovian approximation. This can be checked as follows. Let us introduce the covariance function \eqref{eq:markovian-cov} in Eq.~\eqref{eq:variance-distribution}:
\begin{multline}
\Var{ Q_x} = \int^L_0 \int^L_0 \int^z_0\int^{z'}_0 -\left.\frac{\partial^2}{\partial x^2} A(\boldsymbol{\rho}) \right|_{\boldsymbol{\rho}=0} \delta(s-s') d s\, ds' dz\, dz' \\
= -\left.\frac{\partial^2}{\partial x^2} A(\boldsymbol{\rho}) \right|_{\boldsymbol{\rho}=0}   \int^L_0 ds \int^L_0  dz \Theta(z-s') \int^L_0 dz' \Theta(z' -s)  \\
= -\left.\frac{\partial^2}{\partial x^2} A(\boldsymbol{\rho}) \right|_{\boldsymbol{\rho}=0} \frac{L^3}{3}.
\label{eq:varmark}\end{multline}
It is important to note that the appearance of a third-power law is characteristic from the markovian approximation. 

Since, it is considered an isotropic turbulence the effective covariance function is written as (see Ref.~\cite{book:ishimaru97}, \S{20.2}),
\begin{equation} 
A(\rho) = (2 \pi)^2 \int^{\infty}_0 \kappa d\kappa J_0(\kappa \rho) \Phi(\kappa).
\end{equation}
Thus, substituting the equation above in Eq.~\eqref{eq:varmark} it is exactly reproduced the result obtained by Consortini and O'Donnell \cite{paper:consortini-odonell91} for the long-path case:
\begin{equation}
\Var{Q_x} =\frac{4 \pi^2}{3} L^3 \left\{\int^L_0 \kappa^3 d\kappa \Phi(\kappa) \left[\frac{J_1(\kappa \rho)}{\kappa \rho} - \left(\frac{x}{\rho}\right)^2 J_2(\kappa \rho)\right]\right\}_{\boldsymbol{\rho}=0}.
\end{equation}
It should be stressed that Consortini and O'Donnell obtained this result from Eq.~\eqref{eq:variance-distribution}, so they are within the Geometric Optics approximation. Therefore, it is verified that this approach uses indirectly the markovian approximation. Moreover, it is concluded that the markovian approximation is valid for long-path and weak turbulence as it was mentioned at the end of the introduction. 

On the other hand, if the propagation length $L$ is small the complete variance, Eq.~\eqref{eq:variance-final}, should be calculated. In Ref.~\cite{paper:consortini-odonell91} it is numerically evaluated. The variance is not proportional to the third-power law of the lenght $L$, but asymptotically approaches to it for long-path propagation.

Consortini \etal\, \cite{paper:consortini97} also found a cubic dependence in the strong turbulence case. Having into account that, in this case, the refractive index covariance along the direction of propagation has little effect on the characteristic fluctuation of the ray, the markovian approximation should also be valid for strong turbulence.

Finally, a simple dimensional argument can show the intrinsic third-power law behavior of the markovian approximation. Observe from Eq.~\eqref{eq:markovian-cov} that the refractive index fluctuation behaves as a white noise $W^{1/2}$ along the $z$-axis, so $\nabla_{Q}\epsilon \propto W^{1/2}$. Henceforth, using Eq.~\eqref{eq:small-dis}, the angular beam wandering $\theta$ grows as $L^{1/2}$. The variance of the displacements is proportional to $L^3$.

\section{Fractional Brownian motion and the turbulent refractive index}\label{sec:fbm}

As it is widely known the atmospheric refractive index inherits from the temperature field the quality of being a passive scalar field\footnote{
Remember, that such quantities are diffused and advected by the turbulence with negligible back effect on the flow.
} \cite{book:tatarskii61,book:ishimaru97}---whenever the temperature has also been proved to be a scalar field. Over the last decade the quest of the Fluid Dynamics has been to answer the following question: what properties are passed down by the stochastic nature of the turbulence to the scalars embedded within? The answer is not simple neither has been closed.

Any stochastic variable is characterized through its probability distribution, so the scalars are. First note that it is always possible to assume that any scalar $\vartheta$ is locally homogeneous, just as the velocity field, and therefore its increments are (spatially) stationary random processes---see Ref.~\cite{book:tatarskii61}, p. 19. Thus, it implies that for any statistical moment
\begin{equation*}
\mean{[\vartheta(\vect)-\vartheta(\vect')]^n }=\mean{[\vartheta(\vect-\vect')-\vartheta(0)]^n}, 
\end{equation*}
where $\mean{\,\cdot\,}$ stands for the average over an undefined probability. Nevertheless, it must be stressed that the scalar itself is not stationary; moreover, the gradient of the averaged scalar determines much of its statistical behavior. 

Therefore, it is preferred to study the moments of the increments: the $n$-point structure functions. It has been confirmed that these structure functions vanish if $n$ is odd, and behave as follows otherwise \cite{paper:kraichnan68}:
\begin{equation}
S_{2n}(\vect)=\mean{[\vartheta(\vect)-\vartheta(0)]^{2n}} = A_n \norm{\vect}^{\zeta_{2n}},
\label{eq:n-point-structure-function}\end{equation}
for $\norm{\vect} < L_0$. The coefficients $\zeta_{2n}$ are controlled by the velocity field, an (isotropic) external force acting over the scalar and the gradient of the mean scalar concentration. Given the coefficient $\zeta_2$, a Gaussian behaviour should be observed only when $\zeta_{2n}= n \zeta_2$. Unfortunately, there is plenty of evidence that this is not the case, e.~g.~\cite{paper:shraiman2000}: these exponents deviate from the Gaussian estimated value. Also, the very same coefficient $\zeta_2$ depends on the intermittent behavior of the velocity field itself, and it is different from the `2/3' Kolmogorov exponent.  It has been observed to range from $0$, as the viscosity becomes relevant, to $2$, as the external force reaches the innermost scales~\cite{paper:kraichnan94}. An additive factor called \textit{intermittency exponent} can be introduced to measure these deviations---see for example Ref.~\cite{paper:hentschel83,paper:lvov94,paper:gawedzki95}.

Nevertheless, only the first moments of the propagated light are relevant to us. For that, the  Gaussian distribution should be enough. This is how it has been done while studying scalar turbulence from \textit{synthetic Gaussian velocity fields}---see again references \cite{paper:kraichnan68,paper:shraiman2000,paper:kraichnan94,paper:lvov94,paper:gawedzki95}. 

Finally, the 2-point structure function, Eq.~\eqref{eq:n-point-structure-function} with $n=1$, is called \textit{KOC structure function} or the so called structure function. It is related to the covariance function $v(\vect,\vect')=\langle\vartheta(\vect)\vartheta(\vect')\rangle$:
\begin{equation}
S_2(\vect-\vect') = v(\vect,\vect)+ v(\vect',\vect') - 2 v(\vect,\vect').
\label{eq:struct-cov}
\end{equation}
Observe that the covariance function is symmetric under the change $\vect\leftrightarrow\vect'$; thus, $\partial v(\vect,\vect)/\partial x = \partial v(\vect,\vect)/\partial x'$---the same happens in the other axis. From Eq.~\eqref{eq:struct-cov} it is
\begin{equation*}
\frac{\partial S_2}{\partial x}=2\left[\frac{\partial v}{\partial x}(\vect,\vect)-\frac{\partial v}{\partial x}(\vect,\vect')\right] 
\end{equation*}
Then, from the latter and Eq.~\eqref{eq:n-point-structure-function}, it is obtained:
\begin{equation}
2\left[\frac{\partial v}{\partial x}(\vect,\vect)-\frac{\partial v}{\partial x}(\vect,\vect')\right] = \zeta_2 A_2 \frac{(x - x') }{\norm{\vect-\vect'}^{2-\zeta_2}}
\end{equation}
where $2-\zeta_2>0$; therefore, as $\vect\rightarrow\vect'$ the left-hand side diverges. As it is known a process whose covariance function lacks second partial derivatives is not derivable (see Ref.~\cite{book:cramer67}, \S 9.4); thus, a stochastic process meant to model a scalar should obey this property.

On the other hand, Stolovitzky and Sreenivasan \cite{paper:stolovitzy94} successfully obtained the Kolmogorov's law---the velocity version of Eq.~\eqref{eq:n-point-structure-function}---using \textit{fractional Brownian motion} (fBm) to model the turbulent velocity field.  Therefore, since the phenomenological parallelism between the turbulent velocity field and the scalar fluctuations \cite{paper:shraiman2000} these stochastic processes can also be considered here. It can be seen that they fulfill indeed all the properties mentioned before for the refractive index.

Briefly, fractional Brownian motions are a family of Gaussian processes $B^H$, being $H$ the \textit{Hurst parameter} \cite{paper:hurst65}, with covariance \cite{paper:mandelbrot68}:
\begin{equation}
\E{B^H(t)B^H(s)}=\frac{1}{2}\left(\abs{t}^{2H}+\abs{s}^{2H}-\abs{t-s}^{2H}\right),
\label{eq:fbm-cov}\end{equation}
for $s,t\in\Re$, $0<H<1$, $B^H(0)=0$ almost surely, and $\mathbb{E}[B^H(t)]=0$. Note that these processes are non-differentiable. Also, they are scalar-invariant; that is, $B^H(\alpha s) \overset{d}{=}\alpha^H B^H(s),\text{ for any }\alpha,$ where $\overset{d}{=}$ means both share the same probability law. Usually, scalar-invariant processes are called \textit{self-similar} if they have stationary increments. This is effectively what happens here, as it can be observed from Eq.~\eqref{eq:fbm-cov}. 

One remarkable property of this family $B^H$ is that the $H$ parameter regulates the presence or absence of memory \cite{book:beran94}. In fact, it can be separated in three subfamilies accordingly: long-memory for $1/2<H<1$, no-memory at $H=1/2$, and short-memory in the case $0<H<1/2$.

Now, let the \textit{isotropic fractional Brownian motion} (\textit{i}fBm) be defined as
\begin{equation*}
\tilde{B}^H(\vect) := B^H(\norm{\vect}) = B^H(r).
\end{equation*}
It is straightforward to calculate the covariance of its increments from Eq.~\eqref{eq:fbm-cov}. Afterwards, it is observed that only when $\abs{r-r'}^{3/2}\ll 1$,
\begin{equation}
\E{\left(\tilde{B}^H(\vect) -\tilde{B}^H(\vect') \right)^2} \simeq \norm{\vect - \vect'}^{2H}.
\label{eq:cov-app}\end{equation}

Consequently, the turbulent refractive index can be modeled as follows
\begin{equation}
\epsilon(\vect):= \alpha \tilde{B}^H\!\left(\vect/L_0\right),
\label{eq:fbm-index}
\end{equation}
where $\alpha$ is an adimensional constant. Using the approximation in Eq.~\eqref{eq:cov-app} it is straightforwardly obtained the structure function as below:
\begin{equation*}
\E{(\epsilon(\vect+\vect')-\epsilon(\vect'))^2} = \alpha^2 L_0^{-2H} \norm{\vect}^{2H},
\end{equation*}
for $\norm{\vect}\ll L_0$. Thus, comparing the latter against Eq.~\eqref{eq:n-point-structure-function} for $n=1$, it results $\zeta_2=2H\quad (0<H<1)$ and  $A_2= \alpha^2L_0^{-2H}$---using  typical values of the structure constant for the refractive index and the outer scale $\alpha \sim 10^{-6}$--$10^{-3}$.

Again, a dimensional analysis on the Geometric Optics ray-equation, Eq.~\eqref{eq:small-dis}, can be made to find the path dependence of the refractive index. From Eq.~\eqref{eq:fbm-index}, for $\norm{\mathbf{Q}}\ll 1$, it is: 
\begin{equation*}
\nabla_Q \epsilon = \alpha \frac{W^H\!\!\left(\norm{z\eZ + \mathbf{Q}}\right)}{\norm{z\eZ + \mathbf{Q}}}\, \mathbf{Q}\sim \alpha W^H\!(z)\,\frac{\mathbf{Q}}{z} ,
\end{equation*}
where $W^H$ is the fBm noise. The noise, thought not a derivative in the usual sense, extends the chain rule so the above is valid. Since, $Q/z\sim \theta$ and $\alpha$ is small then $\theta \propto B^H$. Therefore, its variance should grow as $L^{2H}$ and for the displacements as $L^{2H+2}$.

\section{Conclusions}

The procedures employed after the Geometric Optics approximation, which is usually used in atmospheric optics, has been shown equivalent to the markovian approximation. Here is verified its validity in the following cases: at propagation path long enough for weak turbulence, or at any path length for strong tubulence. Moreover, as the cubic dependence on the path length is inherent to the markovian approximation, the Geometric Optics approximations, Eqs. \eqref{eq:non-mark-cov}--\eqref{eq:longpath}, are indirectly based on the former. But, it means that no-memory processes are related with.

A new model for the turbulent refractive index fluctuations is proposed in this paper following most of the requirements of a passive scalar. The isotropic fractional Brownian motion extends the markovian approximation, and introduces memory in the ray-light propagation phenomenon. Furthermore, through the Hurst parameter the state of the scalar turbulence can be set. Then, the variance of the displacements behaves as $L^{2H +2}$. In particular, for Markov processes $H=1/2$, and the cubic dependence is recovered. This is the situation of long path propagation \cite{paper:consortini-odonell91} for weak turbulence, and strong turbulence at any path length \cite{paper:consortini97} discussed in Sec. \ref{sec:markov}.

Finally, the presence of memory, $H\neq 1/2$, was experimentally observed for short path propagation \cite{paper:zunino2004}. Nevertheless, the power-law dependence mentioned above must be experimentally confirmed. That task will be the challenge of future works.

\ack 

DGP has been supported financially by a postdoctoral research fellowship from the Innovatec Foundation (Argentina) and LZ by a doctoral research fellowship from the Consejo Nacional de Investigaciones Cient\'{\i}ficas y T\'ecnicas (CONICET, Argentina). 

The authors are very greateful to the reviewer for constructive criticism.

\bibliography{references}
\bibliographystyle{elsart-num}
\end{document}